\documentclass{ceurart}

\usepackage{listings}
\usepackage{tikz}
\usetikzlibrary{shapes.geometric, arrows.meta, positioning, fit, backgrounds}
\usepackage{booktabs}
\usepackage{caption}
\usepackage{enumitem}
\usepackage{etoolbox}

\setlist{nosep, leftmargin=*, topsep=2pt, partopsep=0pt}
\setlist[itemize]{topsep=2pt, itemsep=1pt}
\setlist[enumerate]{topsep=2pt, itemsep=1pt}
\AtBeginEnvironment{tabular}{\footnotesize}
\makeatletter
\renewcommand{\paragraph}{\@startsection{paragraph}{4}{\z@}%
  {1.5ex \@plus .5ex \@minus .2ex}%
  {-1em}%
  {\normalfont\normalsize\bfseries}}
\renewcommand{\section}{\@startsection{section}{1}{\z@}%
  {-2.5ex \@plus -1ex \@minus -.2ex}%
  {1.5ex \@plus .2ex}%
  {\normalfont\Large\sffamily\bfseries\color{scolor}}}
\renewcommand{\subsection}{\@startsection{subsection}{2}{\z@}%
  {-2ex \@plus -.5ex \@minus -.2ex}%
  {1ex \@plus .2ex}%
  {\normalfont\large\sffamily\bfseries}}
\makeatother

\begin{document}

\ExplSyntaxOn
\cs_set:Npn \__make_tbl_caption:nn #1#2
{
  \l_tbl_align_tl
  \skip_vertical:N \l_tbl_abovecap_skip
  {\parbox{ \dimexpr(\l_tbl_width_dim)}
    {\centering\sffamily\small\textbf{\color{scolor}#1:}~#2\par\vskip2pt }}
  \skip_vertical:N \l_tbl_belowcap_skip
}
\ExplSyntaxOff

\copyrightyear{2026}
\copyrightclause{Copyright for this paper by its authors.
  Use permitted under Creative Commons License Attribution 4.0
  International (CC BY 4.0).}

\conference{}

\title{Semantic Caching for OLAP via LLM-Based Query Canonicalization (Extended Version)}

\author[1]{Laurent Bindschaedler}[%
email=bindsch@mpi-sws.org,
url=https://binds.ch,
orcid=0000-0003-0559-631X,
]
\cormark[1]
\address[1]{Max Planck Institute for Software Systems, Saarbrücken, Germany}

\cortext[1]{Corresponding author.}

\begin{abstract}
Analytical workloads exhibit substantial semantic repetition, yet most production caches key entries by SQL surface form (text or AST), fragmenting reuse across BI tools, notebooks, and NL interfaces. We introduce a safety-first middleware cache for dashboard-style OLAP over star schemas that canonicalizes both SQL and NL into a unified key space---the \emph{OLAP Intent Signature}---capturing measures, grouping levels, filters, and time windows. Reuse requires exact intent matches under strict schema validation and confidence-gated NL acceptance; two correctness-preserving derivations (roll-up, filter-down) extend coverage without approximate matching. Across TPC-DS, SSB, and NYC TLC (1{,}395 queries), we achieve 82\% hit rate versus 28\% (text) and 56\% (AST) with zero false hits; derivations double hit rate on hierarchical queries.
\end{abstract}


\maketitle
\footnotetext{Short version published at DOLAP 2026: 28th International Workshop on Design, Optimization, Languages and Analytical Processing of Big Data, co-located with EDBT/ICDT 2026. This extended version adds detailed background, derivation proofs, expanded evaluation, and discussion.}
\vspace{-24pt}


\section{Introduction}
\label{sec:introduction}

Interactive analytics systems increasingly serve diverse clients: BI dashboards, notebooks, and text-to-SQL interfaces accepting natural language (NL) queries. These use cases amplify latency and compute concerns as LLMs are embedded deeper into analytical pipelines~\cite{mohammadi2025iolmdb}. Query caching is a natural lever because OLAP queries scan large fact tables but return small aggregated results~\cite{gray1997datacube}, making results inexpensive to materialize and reuse.

Analytical caches typically key entries by SQL surface form: normalized query text or Abstract Syntax Tree (AST)-derived canonical representations~\cite{clickhouse2026cache}. NL clients must first translate to SQL, then apply conventional caching. These approaches fragment reuse across heterogeneous clients (Section~\ref{sec:background}): the same OLAP question issued from different tools or phrased differently in NL generates distinct cache keys, causing redundant backend execution.

This paper presents a semantic caching middleware designed for a specific category of queries: dashboard-style OLAP aggregations over star or snowflake schemas with a single fact table and dimension joins. We focus on this well-defined OLAP subset (Section~\ref{subsec:scope}) to make equivalence checkable; queries outside this subset bypass the cache and execute directly on the backend. The main idea is to canonicalize both SQL and NL into a shared, structured cache key, known as an \emph{OLAP Intent Signature}. This signature captures the semantics that affect the numeric result: measures, grouping levels, filters, time windows, and post-aggregation operators. SQL is mapped deterministically from the AST, while NL is translated using a large language model (LLM) constrained to a JSON schema.

The design prioritizes correctness. By default, cache reuse aims for \emph{exact-intent} hits under strict validation. To improve hit rates without relying on approximate matching, we introduce two safe derivations with explicit preconditions: (i) rolling up from finer-grain cached results for additive measures~\cite{park2001olap}, and (ii) filtering down from cached supersets when the cached result includes the necessary filter attributes~\cite{srivastava1996aggregate}. Intent signatures increase reuse by consolidating SQL variants and NL paraphrases into a single key; derivations extend coverage across drill patterns (roll-up, filter tightening).

We evaluate on three decision-support workloads (TPC-DS, SSB, NYC TLC) totaling 1{,}395 queries with systematic SQL variants and NL paraphrases. We compare against (1) normalized text-based caching, (2) AST-based SQL canonicalization, and (3) NL-to-SQL+AST pipelines, measuring hit rate, backend savings, lookup overhead, and false-hit rate. To assess NL canonicalization reliability, we additionally evaluate on adversarial NL queries and BIRD~\cite{li2023bird} human-authored questions.

Across the three primary workloads, our cache achieves 82\% hit rate versus 28.2\% (text) and 55.6\% (AST), reducing backend compute by 85--90\%. Safe derivations raise hit rate from 37\% to 80\% on hierarchical workloads. NL accuracy degrades under ambiguity (44\% adversarial, 51\% BIRD~\cite{li2023bird}), but layered safety prevents incorrect reuse; at threshold 0.5, precision reaches 76.9\% at 36.5\% coverage. Section~\ref{subsec:rq2} details NL risks.

This paper makes the following contributions:
\begin{itemize}
  \item \textbf{Portable intent-signature layer:} An \emph{OLAP Intent Signature} that canonicalizes SQL and NL into a unified key space, enabling cross-client reuse without a single semantic API or platform-specific model.
  \item \textbf{Safety-first reuse with quantified NL failure modes:} Schema validation, confidence gating, and conservative bypassing. We quantify the NL accuracy gap (44\% adversarial, 51\% BIRD~\cite{li2023bird}) and show how safety mechanisms trade coverage for precision.
  \item \textbf{Correctness-preserving derivations:} Roll-up and filter-down from cached results, guarded by explicit preconditions.
\end{itemize}


\section{Background and Motivation}
\label{sec:background}

Query caching is effective for analytical systems~\cite{dar1996semantic,scheuermann1996watchman} because OLAP queries scan large fact tables but return compact aggregated results, making stored results cheap to reuse. However, caching benefits depend on the cache key: semantically identical requests mapped to different keys yield no reuse.

\subsection{Sources of Cache Fragmentation}
Production caches key by SQL surface form---normalized text or AST-derived representations~\cite{clickhouse2026cache}. Three sources fragment reuse in modern analytics.
First, \emph{heterogeneous clients} (BI tools, notebooks, templating systems) generate different SQL for the same question through formatting, alias, and predicate-order variation. AST normalization reduces but does not eliminate this.
Second, \emph{NL interfaces} introduce phrasing diversity without a stable syntactic anchor~\cite{yu2018spider,li2023bird}; even NL-to-SQL translation shifts rather than resolves the problem, since different phrasings yield different SQL.
Third, \emph{approximate semantic matching} (e.g., embedding-based lookups~\cite{bang2023gptcache}) cannot ensure correctness: ``total sales last quarter'' and ``average sales last quarter'' have high similarity but differ by 2--10$\times$ in result.
The net effect: \emph{high semantic repetition does not imply high cache reuse}.

We focus on \emph{query-level} caching (complete results), not cell-level OLAP caches~\cite{mondrian2011cache} that store individual cube cells; our middleware works with any SQL-compatible backend.

\subsection{Motivating Insight}
Dashboard-style OLAP aggregations over star schemas share stable semantic structure~\cite{gray1997datacube}: the numerical output is determined by measures, grouping levels, filters, time windows, and post-aggregation operators. These components are well-defined regardless of whether the query arrives as SQL or NL, and are verifiable against a schema. We key the cache by an \emph{OLAP Intent Signature} encoding these components, collapsing many surface forms into a single semantic key.

Since NL-to-intent mapping can be error-prone, we design the cache around a safety-first pipeline: strict schema validation, confidence-gated reuse, and correctness-preserving derivations (roll-up~\cite{park2001olap}, filter-down~\cite{srivastava1996aggregate}) rather than approximate matching.


\section{System Design}
\label{sec:semantic_cache}

This section describes a middleware cache that answers OLAP-style aggregation queries. It maps each request (SQL or NL) to a structured \emph{OLAP Intent Signature}, the cache key. The design is correctness-first: reuse requires exact intent equality with strict validation. Extensions beyond exact matches are limited to correctness-preserving transformations under stated assumptions.

\subsection{Scope and Assumptions}
\label{subsec:scope}

We focus on a high-value class of dashboard queries, specifically aggregations over a star or snowflake schema~\cite{gray1997datacube} with a single fact table and joins to dimension tables along schema-defined foreign keys. The queries may consist of \texttt{WHERE}, \texttt{GROUP BY}, \texttt{HAVING}, and optionally \texttt{ORDER BY} and \texttt{LIMIT}. We deliberately exclude features that complicate semantic equivalence, such as window functions, set operations, correlated subqueries, recursive CTEs, and lateral joins.

\paragraph{Scope Coverage.} In TPC-DS, 14 of 99 query templates (14\%) qualify; the rest use window functions, CTEs, or set operations outside our scope. SSB (100\%) and NYC TLC (100\%) are fully covered, reflecting their dashboard-oriented design. Queries outside scope execute directly on the backend with no cache interaction, ensuring the system never returns incorrect results for unsupported query patterns.

To make equivalence checkable, we assume that join semantics are dictated by the schema: given the fact table and referenced dimension attributes, there is a unique join path. Requests containing ambiguous join paths (e.g., role-playing dimensions, many-to-many joins, self-joins, or multiple valid paths) are rejected and bypass caching.

\paragraph{Terminology.} We adopt standard OLAP terminology~\cite{golfarelli2009dw}: a \emph{dimension} is a conceptual grouping (e.g., Time, Geography), while a \emph{level} is a specific granularity within a dimension hierarchy (e.g., Year $>$ Quarter $>$ Month). \emph{Roll-up} aggregates from finer to coarser levels; \emph{drill-down} moves from coarser to finer levels. \emph{Slicing} fixes a single dimension value; \emph{dicing} selects ranges across multiple dimensions.

\subsection{Architecture}
\label{subsec:architecture}

\begin{figure}[ht]
\centering
\begin{tikzpicture}[
    node distance=0.3cm and 0.5cm,
    box/.style={rectangle, draw, rounded corners, minimum height=0.55cm, minimum width=1.4cm, font=\footnotesize\sffamily},
    client/.style={box, fill=blue!10, minimum width=0.9cm},
    process/.style={box, fill=orange!15},
    storage/.style={box, fill=green!10},
    output/.style={box, fill=violet!10},
    arrow/.style={-{Stealth[length=1.5mm]}, thick},
    darrow/.style={-{Stealth[length=1.5mm]}, thick, densely dashed, red!60!black},
    lbl/.style={font=\scriptsize\sffamily}
]
\node[client] (sql) {SQL};
\node[client, below=0.15cm of sql] (nl) {NL};

\node[process, right=0.6cm of sql, yshift=-0.25cm] (canon) {Canonicalizer};
\node[process, right=0.5cm of canon] (valid) {Validator};
\node[process, right=0.4cm of valid] (sig) {Hash};
\node[storage, right=0.4cm of sig] (cache) {Cache};
\node[output, right=0.9cm of cache] (result) {Result};

\node[process, below=0.45cm of cache] (backend) {Backend};

\draw[arrow] (sql.east) -- ++(0.15,0) |- (canon.west);
\draw[arrow] (nl.east) -- ++(0.15,0) |- (canon.west);
\draw[arrow] (canon) -- (valid);
\draw[arrow] (valid) -- (sig);
\draw[arrow] (sig) -- (cache);
\draw[arrow] (cache) -- node[above, lbl] {hit} (result);

\draw[arrow] ([xshift=2mm]cache.south west) -- node[left, lbl] {miss} ([xshift=2mm]backend.north west);
\draw[arrow] ([xshift=-2mm]backend.north east) -- node[right, lbl] {populate} ([xshift=-2mm]cache.south east);
\draw[arrow] (backend.east) -| node[below, lbl, pos=0.3] {execute} (result.south);

\draw[darrow] (valid.south) |- node[lbl, pos=0.75, below] {bypass} (backend.west);

\end{tikzpicture}
\caption{Architecture: queries are canonicalized, validated, and hashed as cache keys. Hits return cached results; misses execute on backend. Validation failures bypass the cache (dashed path). Derivation checks (roll-up, filter-down) occur within cache lookup (not shown).}
\label{fig:architecture}
\end{figure}
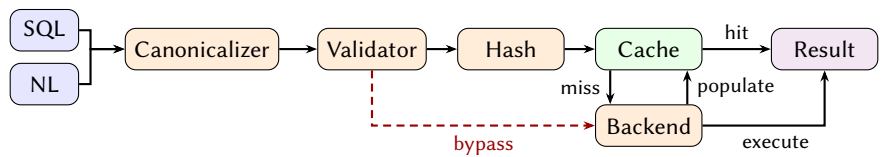

Figure~\ref{fig:architecture} illustrates the system architecture. The middleware operates as an intermediary between clients, such as BI tools, notebooks, and NL interfaces, and the backend OLAP engine.

For each request, the following steps are taken: (1) canonicalize the request into an intent signature, (2) validate the signature against the schema and safety rules, (3) look up the signature hash in the cache, (4) on a miss, execute on the backend and store the result under the signature. Because canonicalization and validation are distinct steps, reuse decisions are auditable: the signature is an explicit contract that determines whether a cached result may be returned.

\subsection{OLAP Intent Signature}
\label{subsec:signature}

We define a canonical signature as a JSON object containing all semantics that can affect the numerical output of the query. The components of the signature include:
\begin{itemize}
  \item \textbf{Measures}: aggregation function and base expression (for example \texttt{SUM(f\_sales)}), including \texttt{DISTINCT} flags.
  \item \textbf{Grouping levels}: list of hierarchy levels in \texttt{GROUP BY}, sorted alphabetically to a canonical order (for example \texttt{geo.region}, \texttt{time.month}). SQL GROUP BY order is semantically irrelevant; we canonicalize to ensure equivalent queries produce identical signatures.
  \item \textbf{Filters}: a set of normalized predicates over non-temporal dimensions and facts, including canonical literal formats.
  \item \textbf{Time window}: explicit start and end boundaries on the time dimension, normalized to a canonical representation. We separate time windows from general filters for two reasons: (i) temporal predicates require special canonicalization (resolving ``last quarter'' to concrete dates, handling timezone normalization), and (ii) time boundaries are critical for cache invalidation---when new data arrives, entries with open-ended time windows (``last 30 days'') must be refreshed while closed windows (``Q1 2024'') remain valid.
  \item \textbf{Post-aggregation operators}: \texttt{HAVING}, \texttt{ORDER BY}, \texttt{LIMIT}.
  \item \textbf{Metric identity (optional)}: a metric identifier from a governed layer, if available.
  \item \textbf{Scope (optional)}: for multi-tenant deployments, a tenant or user identifier ensures cache isolation across security boundaries.
\end{itemize}
The optional fields are relevant for governed or multi-tenant deployments; our evaluation uses ungoverned single-tenant schemas.

We serialize the signature into a canonical JSON string (with sorted keys and normalized lists) and compute a hash (SHA-256) to obtain a fixed-length cache key. This ensures different surface forms map to the same signature (Figure~\ref{fig:signature_example}).

\paragraph{Query Model and Join Handling.} Join paths are implicit---determined by the schema's foreign keys---and not included in the signature. The canonicalizer verifies that all referenced columns exist and join paths are unambiguous. Queries with self-joins, role-playing dimensions, or many-to-many relationships bypass the cache, ensuring that matching signatures produce identical results.

\begin{figure}[ht]
\centering
\fbox{\begin{tabular}{@{}p{0.46\columnwidth}@{\hspace{0.04\columnwidth}}p{0.46\columnwidth}@{}}
\multicolumn{2}{@{}l@{}}{\footnotesize\textbf{NL:} ``Show total revenue by region for electronics in Q1 2024''} \\[0.5em]
\footnotesize\textbf{SQL:}\newline
{\ttfamily\footnotesize
SELECT r.region\_name,\newline
\hphantom{SELEC}SUM(s.amount) AS revenue\newline
FROM sales s\newline
\hphantom{FR}JOIN regions r ON s.region\_id = r.id\newline
\hphantom{FR}JOIN products p ON s.product\_id = p.id\newline
WHERE p.category = 'electronics'\newline
\hphantom{WH}AND s.sale\_date >= '2024-01-01'\newline
\hphantom{WH}AND s.sale\_date < '2024-04-01'\newline
GROUP BY r.region\_name}
&
\footnotesize\textbf{OLAP Intent Signature:}\newline
{\ttfamily\footnotesize
\{\newline
\hphantom{XX}"measures": [\{"agg": "SUM",\newline
\hphantom{XXXX}"expr": "sales.amount"\}],\newline
\hphantom{XX}"levels": ["regions.region\_name"],\newline
\hphantom{XX}"filters": [\{"col": "products.category",\newline
\hphantom{XXXX}"op": "=", "val": "electronics"\}],\newline
\hphantom{XX}"time\_window": \{\newline
\hphantom{XXXX}"start": "2024-01-01",\newline
\hphantom{XXXX}"end": "2024-04-01"\}\newline
\}}
\\
\end{tabular}}
\caption{Example: an NL question and equivalent SQL query produce identical OLAP Intent Signatures.}
\label{fig:signature_example}
\end{figure}

\subsection{Canonicalization}
\label{subsec:canonicalization}

SQL requests are parsed into an AST and normalized deterministically (identifier resolution, predicate ordering, literal canonicalization). NL requests are mapped to signatures using an LLM constrained to produce strict JSON; the LLM also returns a confidence score (0--1) used for safety gating. Low-confidence signatures bypass caching to avoid incorrect reuse.

\paragraph{SQL $\rightarrow$ Signature.}
The AST is normalized deterministically (identifier resolution, commutative predicate ordering, literal canonicalization) and intent components extracted. Identical signatures imply identical semantics under the schema conditions in Section~\ref{subsec:scope}.

\paragraph{NL $\rightarrow$ Signature.}
The LLM prompt includes the schema and a controlled vocabulary of measures and dimensions. The returned confidence score (0--1) is an uncalibrated heuristic used for coarse gating (Section~\ref{subsec:safety_policy}); more principled alternatives (calibration, log-probabilities, self-consistency) are deferred to future work. Even simple gating substantially improves precision (Table~\ref{tab:safety_mechanisms}).

\subsection{Validation and Cache Lookup}
\label{subsec:validation_lookup}

Before reuse, we validate that all referenced measures/dimensions exist, time windows resolve to concrete boundaries, and join paths are unambiguous. Validation failures bypass the cache and execute on the backend, prioritizing misses over incorrect reuse.

Before reusing any elements, we perform the following validations:
(1) all referenced measures/dimensions exist and pass type checks,
(2) the specified time window resolves to concrete boundaries,
(3) the implied join path is unique within the schema, and
(4) unsupported constructs trigger bypass.

If any of these validations fail, the request will still execute on the backend. However, it may only be stored under the \emph{executed} signature if the signature is well-formed and the policy allows it. This approach enforces a conservative default: prioritize misses over incorrect reuse. When the cache reaches capacity, entries are evicted using a Least Recently Used (LRU) policy, which discards the entry that has gone longest without being accessed.

\subsection{Correctness-Preserving Reuse Beyond Exact Hits}
\label{subsec:derivations}

Exact intent matching is the default reuse mode. We also support two safe derivations:

\paragraph{Roll-Up.} Re-aggregation from a finer-grained cached entry is permitted only for composable aggregations~\cite{gray1997datacube,lenz1997summarizability,pedersen1999preaggregation} (\texttt{SUM}, \texttt{COUNT}, \texttt{MIN}, \texttt{MAX}); non-additive measures (\texttt{AVG}, \texttt{COUNT DISTINCT}, ratios) are rejected. Preconditions: (i) composable aggregation, (ii) summarizable hierarchy (each child maps to exactly one parent at each level), (iii) NULL-preserving semantics.

\paragraph{Filter-Down.} The cached result must contain the filter attributes needed for the tighter predicate; otherwise the derivation is rejected. Preconditions: (i) filter attributes present in cached columns, (ii) SQL NULL semantics preserved, (iii) no \texttt{ORDER BY}/\texttt{LIMIT}.

Derivations are disabled under \texttt{ORDER BY} or \texttt{LIMIT} because re-aggregation or post-filtering can alter top-k membership. Drill-down (finer $\leftarrow$ coarser) is unsupported: query-level caching lacks the detail data. Drill-down queries populate the cache but do not reuse coarser entries.

\subsection{Safety Policy for NL-Driven Reuse}
\label{subsec:safety_policy}

NL canonicalization can be schema-valid yet semantically incorrect (e.g., mapping an ambiguous term to the wrong column). We control NL reuse via layered policies:
\begin{itemize}
  \item \textbf{Schema validation:} required for all reuse to ensure structural correctness.
  \item \textbf{Confidence-gated reuse:} low-confidence signatures bypass caching.
  \item \textbf{Heuristic ambiguity checks:} reject common ambiguity patterns, such as unresolved relative time, underspecified spatial terms, and aggregation-word mismatches.
  \item \textbf{Optional lightweight verification:} a configurable check on NL-originated hits that can catch some classes of mismatch, primarily time-window errors.
\end{itemize}

These layers are designed to prioritize misses over false hits, ensuring that the reuse behavior aligns with the requirements for analytics correctness.


\section{Prototype Implementation}
\label{sec:implementation}

We implement the middleware in Python. The \emph{canonicalizer} parses SQL with \texttt{sqlglot} and normalizes deterministically (identifier resolution, commutative predicate ordering, literal canonicalization); NL requests are mapped via an LLM endpoint emitting schema-constrained JSON with confidence scores. The \emph{validator} checks measure/dimension existence, time-window resolution, and join-path uniqueness; failures bypass the cache.

The \emph{cache store} persists results as Parquet files indexed by signature hash, with a SQLite metadata index for derivation candidate lookup (entries matching requested measures with superset dimensions or superset filters). The executor uses DuckDB as backend. NL-string $\rightarrow$ signature mappings are memoized to avoid repeat LLM calls. Configuration knobs include confidence threshold, heuristic toggles, derivation enable/disable, and snapshot strategy for invalidation.

\paragraph{Reproducibility.} We will release source code, the LLM prompt template ($\sim$800 tokens), workload generation scripts, the JSON Schema specification, and the 63 adversarial queries with annotations.


\section{Evaluation}
\label{sec:evaluation}

We evaluate the intent-signature cache along four research questions: \textbf{RQ1} hit rate, cache fragmentation, and NL reuse patterns, \textbf{RQ2} correctness (false hits), \textbf{RQ3} backend savings and overheads, and \textbf{RQ4} coverage gains from correctness-preserving derivations.

\subsection{Experimental Setup}
\label{subsec:experimental_setup}

\paragraph{Workloads.}
We use \textbf{TPC-DS}~\cite{tpcds} (SF=1, 14 in-scope queries), \textbf{SSB}~\cite{oneil2009ssb} (13 queries), and \textbf{NYC TLC}~\cite{nyctlc} (18 templates): 1{,}395 queries total (945 SQL, 450 NL). For NL robustness, we use 63 \textbf{adversarial NL queries} stressing ambiguity and 150 \textbf{BIRD}~\cite{li2023bird} human-authored OLAP-compatible questions (9.8\% of BIRD's dev set, selected per Section~\ref{subsec:scope}). For derivation evaluation, we construct an \textbf{SSB hierarchical workload} drilling through time, geography, and product hierarchies.

\paragraph{Query Variants and Ground Truth.}
Per canonical intent, we generate 21 SQL variants (formatting, alias, predicate-order changes) and 10 manually authored NL paraphrases. All variants are verified to produce identical results, establishing ground truth for hit-rate and false-hit measurement.

\paragraph{Baselines.}
We compare: (1)~\textbf{TextCache} (normalized SQL text), (2)~\textbf{ASTCache} (AST-derived canonicalization), (3)~\textbf{NL-to-SQL+AST} (NL translated to SQL, then ASTCache), (4)~\textbf{LLMSigCache} (our system). TextCache/ASTCache are SQL-only.

\paragraph{Platform.}
Server: two AMD EPYC 9654 CPUs, 2\,TB DDR5 RAM, Debian Linux. Backend: DuckDB. NL uses GPT-4o-mini at temperature~0; Table~\ref{tab:config_ablation}b reports model comparison. Three trials per method; NL memoization cleared between trials.

\subsection{RQ1: Does Intent-Signature Caching Improve Hit Rates?}
\label{subsec:rq1}

Table~\ref{tab:main_results} reports hit rates and reduction factors. LLMSigCache achieves the highest hit rate on all three workloads (82\% average), outperforming TextCache (28.2\%) and ASTCache (55.6\%). Relative to ASTCache, the gain is driven by NL: ASTCache cannot process NL inputs, while LLMSigCache maps NL paraphrases into the same intent key space and enables reuse.

\begin{table}[ht]
\centering
\caption{Cache performance by method. Hit rate (\%); Reduction factor (queries per cache key, higher = less fragmentation). TextCache/ASTCache reduction factors are SQL-only (68\% of workload).$^\dagger$}
\label{tab:main_results}
\begin{tabular}{l@{\hspace{1em}}ccc@{\hspace{0.8em}}c@{\hspace{1.2em}}ccc}
\toprule
 & \multicolumn{4}{c}{\textbf{Hit Rate (\%)}} & \multicolumn{3}{c}{\textbf{Reduction ($\times$)}} \\
\cmidrule(lr){2-5} \cmidrule(lr){6-8}
\textbf{Method} & \textbf{NYC TLC} & \textbf{SSB} & \textbf{TPC-DS} & \textbf{Avg} & \textbf{NYC TLC} & \textbf{SSB} & \textbf{TPC-DS} \\
\midrule
TextCache      & 13.8 & 38.2 & 32.7 & 28.2 & 2.1 & 3.7 & 3.2 \\
ASTCache       & 63.4 & 54.1 & 49.3 & 55.6 & 32.7 & 31.0 & 24.1 \\
NL-to-SQL+AST  & 86.6 & 80.4 & 65.0 & 77.3 & 9.4 & 16.1 & 8.0 \\
LLMSigCache    & 96.9 & 81.9 & 67.1 & 82.0 & 19.0 & 16.8 & 9.0 \\
\bottomrule
\end{tabular}

{\footnotesize $^\dagger$ASTCache excludes NL queries (32\% of workload), inflating its per-key ratio on the SQL subset.}
\end{table}

Within LLMSigCache, NL reuse arises from both \emph{NL-to-NL} hits (paraphrases of the same intent) and \emph{cross-surface} hits (NL matching SQL-populated entries or vice versa). Cross-surface sharing accounts for 3--34\% of NL cache hits; most NL benefit comes from unifying paraphrases under intent keys.

All methods produce \textbf{zero false hits} on controlled workloads. NL accuracy drops to 44\% on adversarial queries and 51\% on BIRD~\cite{li2023bird} human-authored questions, making safety essential: a 0.5 confidence threshold yields 77\% precision at 37\% coverage. Backend compute drops 85--90\%. Safe derivations (roll-up, filter-down) raise hit rate from 37\% to 80\% on hierarchical workloads.

\subsection{RQ2: Does Intent-Signature Caching Preserve Correctness?}
\label{subsec:rq2}

We distinguish \emph{cache correctness} (a hit returns the same result as backend execution) from \emph{user-intent correctness} (the signature matches what the user meant). Zero false hits ensure cache correctness; NL accuracy determines user-intent correctness.

\paragraph{False Hits for Main Workloads.}
Across all 1{,}395 controlled queries, all methods exhibit \textbf{zero false hits}. Signature specificity (5+ structured fields, SHA-256 hashing) makes collisions rare; SQL-seeded reuse (Section~\ref{subsec:deployment_knobs}) restricts NL to read-only cache access for stronger guarantees.

\paragraph{NL Semantic Accuracy under Ambiguity.}
Cache correctness depends on NL canonicalization accuracy when NL requests are allowed to reuse cached results. Therefore, we evaluate canonicalization under two stress settings.

First, we construct 63 adversarial NL queries designed to trigger ambiguity (metric name, time references, dimension disambiguation, aggregation intent, and compositional requests). Table~\ref{tab:semantic_accuracy} shows that exact semantic accuracy is 44.4\% on these adversarial queries, with many errors being \emph{schema-valid but semantically wrong}.

\begin{table}[ht]
\centering
\caption{Semantic accuracy on 63 adversarial NL queries by ambiguity type.}
\label{tab:semantic_accuracy}
\begin{tabular}{llcccc}
\toprule
\textbf{Ambiguity Type} & \textbf{Example} & \textbf{N} & \textbf{Correct} & \textbf{Wrong} & \textbf{Invalid} \\
\midrule
Metric name    & ``revenue'' $\rightarrow$ \texttt{net\_amount} vs \texttt{gross\_amount} & 15 & 11 & 4 & 0 \\
Time reference & ``last month'' without current date context & 12 & 2 & 10 & 0 \\
Dimension      & ``area'' $\rightarrow$ \texttt{zone} vs \texttt{borough} & 12 & 3 & 9 & 0 \\
Aggregation    & ``average trips'' $\rightarrow$ AVG vs COUNT & 9 & 6 & 3 & 0 \\
Compositional  & Multiple measures in single query & 15 & 6 & 4 & 5 \\
\midrule
\textbf{Total} & --- & 63 & 28 & 30 & 5 \\
\bottomrule
\end{tabular}
\end{table}

Second, we evaluated 150 human-authored OLAP-compatible questions from BIRD~\cite{li2023bird}, using the corresponding schemas. The resulting semantic accuracy is 51.3\% in this setting. The main workload paraphrases are controlled rewrites of known-correct intents with unambiguous references; BIRD and adversarial queries introduce realistic ambiguity (synonyms, implicit time references, underspecified dimensions) that is absent from the main evaluation, explaining the accuracy gap.

These results quantify a practical gap: NL canonicalization is the limiting factor under realistic ambiguity. They motivate the safety policy in Section~\ref{subsec:safety_policy}, which trades coverage for precision via confidence-gated reuse and heuristic ambiguity checks.

Table~\ref{tab:safety_mechanisms}a reports the coverage--precision trade-off: at threshold 0.5, precision rises to 76.9\% at 36.5\% coverage. Table~\ref{tab:safety_mechanisms}b shows schema-specific heuristics that reject common ambiguity patterns (unresolved relative time, underspecified spatial terms, aggregation-word mismatches). Heuristics reduce wrong signatures from 30 to 9, improving precision from 48\% to 69\%, but reject 54\% of queries. These are deployment-specific templates; operators define analogous rules per schema.

\begin{table}[ht]
\centering
\caption{Safety mechanism effectiveness on adversarial queries (N=63).}
\label{tab:safety_mechanisms}
\begin{minipage}[t]{0.48\linewidth}
\centering
\textbf{(a) Confidence threshold}
\vspace{2pt}

\begin{tabular}{lcc}
\toprule
\textbf{Threshold} & \textbf{Coverage} & \textbf{Precision} \\
\midrule
0.3 & 63.5\% & 62.5\% \\
0.5 & 36.5\% & 76.9\% \\
0.7 & 20.6\% & 78.3\% \\
0.9 & 12.7\% & 100.0\% \\
\bottomrule
\end{tabular}
\end{minipage}%
\hfill
\begin{minipage}[t]{0.48\linewidth}
\centering
\textbf{(b) Schema-specific heuristics}
\vspace{2pt}

\begin{tabular}{lcc}
\toprule
 & \textbf{Validation only} & \textbf{With heuristics} \\
\midrule
Precision        & 48.3\% & 69.0\% \\
Wrong signatures & 30     & 9 \\
Bypass rate      & 7.9\%  & 54.0\% \\
\bottomrule
\end{tabular}
\end{minipage}
\end{table}

\subsection{RQ3: What Are the Backend Savings and Overheads?}
\label{subsec:rq3}

\paragraph{Backend Savings.}
LLMSigCache reduces backend compute by 85--90\%, comparable to ASTCache on the SQL subset and higher overall because it also caches NL.

\paragraph{Overhead.}
SQL lookup adds 9--16\,ms median (Table~\ref{tab:latency}). NL pays $\sim$1.3\,s LLM cost on first occurrence; exact repeats are memoized. SQL hits save 150--800\,ms backend time at 10--20\,ms lookup cost. Derivations operate on in-memory Parquet and remain sub-second.

\paragraph{Capacity Sensitivity.}
Table~\ref{tab:capacity} shows LRU eviction under varying cache sizes. Dashboard-like orderings (Sequential, Zipf) remain effective at 10--25\% capacity; Interleaved requires larger caches.

\begin{table}[ht]
\centering
\caption{RQ3 overhead measurements.}
\label{tab:latency}\label{tab:capacity}

\begin{minipage}[t]{0.42\linewidth}
\centering
\textbf{(a) Latency (ms) by scenario}
\vspace{2pt}

\begin{tabular}{lcc}
\toprule
\textbf{Scenario} & \textbf{Med.} & \textbf{P95} \\
\midrule
\multicolumn{3}{l}{\textit{SQL queries (all methods)}} \\
Cache lookup      & 9--16 & 12--28 \\
\midrule
\multicolumn{3}{l}{\textit{NL queries (LLMSigCache)}} \\
First occur. (miss)  & 1{,}317 & 2{,}266 \\
First occur. (hit)   & 1{,}304 & 2{,}016 \\
Repeat (memo)        & $<$0.01 & $<$0.01 \\
\bottomrule
\end{tabular}
\end{minipage}%
\hfill
\begin{minipage}[t]{0.55\linewidth}
\centering
\textbf{(b) Hit rate (\%) vs.\ cache size (NYC TLC)}
\vspace{2pt}

\begin{tabular}{lccccc}
\toprule
\textbf{Ordering} & \textbf{10\%} & \textbf{25\%} & \textbf{50\%} & \textbf{75\%} & \textbf{100\%} \\
\midrule
Sequential   & 96.8 & 96.8 & 96.8 & 96.8 & 96.9 \\
Random       & 4.9  & 22.9 & 47.5 & 69.9 & 96.9 \\
Interleaved  & 0.0  & 5.1  & 5.1  & 5.1  & 96.9 \\
Zipf         & 13.4 & 45.2 & 72.8 & 85.7 & 96.9 \\
\bottomrule
\end{tabular}
\end{minipage}
\end{table}

\subsection{RQ4: Do Correctness-Preserving Derivations Extend Coverage?}
\label{subsec:rq4}

To evaluate derivations, we construct an SSB hierarchical workload where queries drill up and down through time, geography, and product hierarchies. SSB's explicit dimension hierarchies are representative of dashboard drill patterns; TPC-DS and NYC TLC lack systematic hierarchy traversal, so derivations provide limited benefit on those workloads by design. Without derivations, hit rate is 37\%; enabling roll-up and filter-down raises it to 80\% with zero false hits.


\section{Discussion and Deployment Considerations}
\label{sec:discussion}

\subsection{Deployment Knobs}
\label{subsec:deployment_knobs}

The main deployment risk is NL canonicalization producing \emph{schema-valid but semantically incorrect} signatures. With only 44\% accuracy on adversarial queries (Table~\ref{tab:semantic_accuracy}), confidence-gated reuse and ambiguity heuristics are essential. Table~\ref{tab:config_ablation} summarizes configuration profiles controlling the precision-coverage trade-off. Additional restrictions include \emph{SQL-seeded reuse only} (preventing cache poisoning from NL) and \emph{scoped signatures} with tenant context. In governed deployments (e.g., dbt Metrics, Cube), metric identifiers in the signature can eliminate ambiguity at the source.

\begin{table}[ht]
\centering
\caption{Configuration profiles and model comparison on adversarial queries (N=63).}
\label{tab:config_ablation}

\textbf{(a) Configuration profiles}
\vspace{2pt}

\begin{tabular}{@{}lccc@{}}
\toprule
\textbf{Setting} & \textbf{Conservative} & \textbf{Balanced} & \textbf{Aggressive} \\
\midrule
Schema validation & Yes & Yes & Yes \\
Confidence threshold & 0.7 & 0.5 & None \\
Schema heuristics & All & Time+spatial & None \\
\midrule
NL precision     & 71.4\% & 72.0\% & 48.3\% \\
NL coverage      & 22.2\% & 39.7\% & 92.1\% \\
Wrong cached     & 4 & 7 & 30 \\
\bottomrule
\end{tabular}

\vspace{8pt}
\textbf{(b) LLM ablation}
\vspace{2pt}

\begin{tabular}{@{}lcccc@{}}
\toprule
\textbf{Model} & \textbf{Correct} & \textbf{Wrong} & \textbf{Invalid} & \textbf{Accuracy} \\
\midrule
GPT-4o-mini      & 28 & 30 & 5 & 44.4\% \\
Claude-3.5-haiku & 38 & 25 & 0  & 60.3\% \\
\bottomrule
\end{tabular}
\end{table}

\subsection{Cache Invalidation}
\label{subsec:invalidation_discussion}

Our evaluation uses static snapshots, but production systems must link entries to data freshness~\cite{gupta1995viewmaintenance}. The prototype records snapshot identifiers per entry; entries are invalidated on schema change or when their time window intersects updated partitions. For append-only fact tables, closed time windows (``Q1 2024'') remain valid indefinitely; only open-ended windows (``last 30 days'') require refresh. Empirical characterization of cache churn under realistic update patterns---including closed-vs.-open window ratios, update frequency, and dimension change rates---remains future work.

\subsection{Limitations}
\label{subsec:limitations}

\paragraph{Restricted Query Class.}
Queries with window functions, set operations, or CTEs bypass the cache. Derived metrics (ratios, year-over-year) require expression-tree signatures or metric-layer integration.

\paragraph{NL Canonicalization Quality.}
NL accuracy drops on ambiguous inputs (Table~\ref{tab:semantic_accuracy}); safety policies trade coverage for precision.

\paragraph{Derivation Limits.}
Roll-up requires composable aggregations and is disabled for non-additive measures and ordering/top-k queries.

\paragraph{Synthetic Workload.}
Controlled variants may overstate hit rates vs.\ production traffic. Alternative generators such as CubeLoad~\cite{rizzi2014cubeload} and production trace validation would complement controlled benchmarks.


\section{Related Work}
\label{sec:related_work}

\paragraph{Semantic Caching and Decision-Support Reuse.}
Classic semantic caching stores results using semantic descriptions (e.g., predicate regions) and reuses them when a new query can be answered from cached data~\cite{dar1996semantic}. WATCHMAN frames caching as a workload-aware manager for decision-support queries~\cite{scheuermann1996watchman}. We follow the same goal but focus on a restricted \emph{intent signature} that serves as a deterministic cache key derivable from both SQL and NL.

\paragraph{Answering Aggregates Using Views.}
Prior work addresses answering aggregation queries from previously computed views~\cite{srivastava1996aggregate,kotidis1999dynamat,goldstein2001views} and rewriting using dimension hierarchies~\cite{park2001olap} under summarizability constraints~\cite{lenz1997summarizability,pedersen1999preaggregation}. Our derivations are correctness-preserving rewrites guarded by explicit preconditions. Unlike materialized view selection, which optimizes offline view sets, we perform runtime caching with online key derivation from both SQL and NL.

\paragraph{SQL and NL Query Caching.}
Many systems use normalized SQL text or AST-based canonical forms to key caches~\cite{clickhouse2026cache}. Text-to-SQL research~\cite{yu2018spider,li2023bird,pourreza2023dinsql,chaturvedi2025sqlt} enables NL$\rightarrow$SQL$\rightarrow$cache-key pipelines. VOOL~\cite{francia2025vool} provides a modular framework for vocalizing OLAP sessions, mapping between NL and OLAP operations; our work shares the goal of bridging NL and structured OLAP semantics but focuses specifically on caching with correctness guarantees rather than session vocalization. We key by OLAP semantic contract directly, extending coverage to both SQL variants and NL paraphrases.

\paragraph{BI Semantic Layer Caches.}
Enterprise BI servers (e.g., Oracle BI Server, Looker, Cube~\cite{CubeSemanticLayer}, dbt Semantic Layer~\cite{dbtSemanticLayer}) implement logical caches with subset and roll-up reuse rules similar to our derivations. However, these caches are keyed by platform-specific logical models; cache reuse requires queries to be routed through the platform's query engine. Our contribution is a \emph{portable middleware} that canonicalizes both heterogeneous SQL and NL into a unified semantic key space for caching, enabling cross-client and cross-modality reuse without requiring adoption of a single platform.

\paragraph{Approximate Matching and Embedding Caches.}
Embedding-based caches use semantic similarity but cannot distinguish ``total revenue'' from ``average revenue''~\cite{bang2023gptcache}; our zero-tolerance for false hits makes them unsuitable. General query equivalence is hard~\cite{chandra1977containment,klug1982aggregate}; our intent signature provides a practical fingerprint for a restricted but common fragment.


\section{Conclusion and Future Work}
\label{sec:conclusion}

We introduced a safety-first semantic caching middleware for dashboard-style OLAP, keyed by an \emph{OLAP Intent Signature} encoding measures, grouping levels, filters, time windows, and post-aggregation operators. Intent-signature caching achieves 82\% hit rate across three workloads (vs.\ 28.2\% text, 55.6\% AST), yielding 85--90\% backend savings; derivations raise hierarchical hit rates from 37\% to 80\%. NL canonicalization accuracy drops on adversarial inputs, necessitating conservative safety policies. Unlike BI semantic layers, our approach unifies heterogeneous SQL and NL into a single, portable cacheable representation.

\paragraph{Future Work.}
Robust invalidation under data updates is the key gap for production deployment. Promising directions include: stronger schema grounding for NL canonicalization; richer cube-lattice derivations~\cite{harinarayan1996datacube}; validation against production query traces; and extending input modalities to MDX (the multidimensional query language for OLAP cubes), which would require an MDX parser but shares the same underlying intent structure.

\section*{Declaration on Generative AI}
The LLMs evaluated in this work (GPT-4o-mini, Claude-3.5-haiku) are components of the proposed system and were not used to prepare the manuscript. The author used Claude (Anthropic) and Grammarly for writing style, grammar, spelling, and formatting, and OpenAI Deep Research for citation management. The author reviewed and edited all outputs and takes full responsibility for the content.

\bibliography{main}

\end{document}